\documentclass[twocolumn]{article} 

\usepackage{kotex} 
\usepackage{url}
\usepackage{amsmath} 
\usepackage{graphicx} 
\usepackage{xcolor}
\usepackage{hyperref}
\usepackage[margin=2cm]{geometry}

\usepackage{siunitx}
\usepackage{chemformula}
\DeclareSIUnit{\molar}{M}
\DeclareGraphicsExtensions{.pdf,.eps,.png,.jpg,.mps} 
\title{On-chip label-free biosensing based on active whispering gallery mode resonators pumped by a light-emitting diode}
\author{Yeseul Kim$^1$ and Hansuek Lee$^{1,2}$*}
\date{\small
 $^1$Department of Physics, Korea Advanced Institute of Science and Technology, Daejeon 34141, Republic of Korea\\
 $^2$Graduate School of Nanoscience and Technology, Korea Advanced Institute of Science and Technology, Daejeon 34141, Republic of Korea\\
 \url{hansuek@kaist.ac.kr}}

\begin{document}

\twocolumn[
\begin{@twocolumnfalse}
	\maketitle
	\begin{abstract}

	Biosensing based on whispering-gallery mode (WGM) resonators has been continuously studied with great attention due to its excellent sensitivity guaranteeing the label-free detection. However, its practical impact is insignificant to date despite notable achievements in academic research. Here, we demonstrate a novel practical platform of on-chip WGM sensors integrated with microfluidic channels. By placing silicon nanoclusters as a stable active compound in micro-resonators, the sensor chip can be operated with a remote pump and readout, which simplifies the chip integration and connection to the external setup. In addition, silicon nanoclusters having large absorption cross-section over broad wavelength range allow active sensing for the first time with an LED pump in a top-illumination scheme which significantly reduces the complexity and cost of the measurement setup. The nano-slot structure of 25 $\si{\nano\meter}$ gap width is embedded in the resonator where the target bio-molecules are selectively detected with the sensitivity enhanced by strongly confined mode-field. The sensitivity confirmed by real-time measurements for the streptavidin-biotin complex is 0.012 $\si{\nano\meter/ \nano\molar}$, improved over 20 times larger than the previously reported WGM sensors with remote readout.

	\vspace{0.3cm}
		
	\end{abstract}
\vspace{2cm}
\end{@twocolumnfalse}
]


\newpage
\section{Introduction}

Sensing based on WGM resonators has been considered as one of the most promising approaches to realize lab-on-a-Chip (LoC) devices for label-free detection of bio/chemical molecules and particles.\cite{Eugene2017towards,vollmer2008whispering,vollmer2012review, subramanian2018label}
Since the sensing event is observed by the frequency shift of WGM resonance caused by the fractional perturbation of mode volume, any kinds of particles having the refractive index different from that of the environment can be detected without the need for the labeling.\cite{Eugene2017towards, baaske2014single}
This versatile method providing outstanding sensitivity by means of high quality factor and small mode volume of WGM resonators has been verified by the detection of various kinds of nanoparticles including polystyrene beads, viruses, and DNA.\cite{vollmer2008whispering,he2011detecting, armani2007label, zhu2010chip, arnold2010whispering, foreman2015whispering}
In addition, it has also demonstrated great potential for the analysis of molecular dynamics based on the real-time measurement of specific biomolecule interactions.\cite{vollmer2012review,foreman2015whispering}
Since WGM resonators can be fabricated on a chip in micrometer-scale footprint, it has been expected that this approach can be implemented in a form of miniaturized devices integrated on a chip.\cite{Eugene2017towards, wienhold2015all, xu2018wireless}
However, despite these benefits, most of the WGM-based sensing researches primarily remain in scientific interest apart from the development of practical LoC devices.\cite{Eugene2017towards}

The fundamental restriction hindering the development of the practical devices mainly comes from the light coupling scheme based on evanescent coupling which requires a waveguide physically connecting between the resonator and the external setup. 
For example, tapered optical fibers which are the most commonly used waveguides for the coupling,\cite{knight1997phase} are mechanically unstable and difficult to combine with microfluidic channels.\cite{wienhold2015all}
It also needs a precise method to adjust the gap between the waveguide and resonator in nanometer-scale to control coupling efficiency. On the other hand, bus waveguides monolithically implemented on a resonator chip are fairly robust but requires high microfabrication precision to define the gap and additional effort to implement light coupling at the end of the waveguides such as an end-fire coupling.\cite{barrios2012integrated}

To overcome this limitation, the concept of the WGM sensor based on optically active resonators has emerged for promising alternatives.\cite{Eugene2017towards} In this approach, pump light focused to the active resonators induces the emission spectrum peaks along with the resonance modes which are detected by a spectrometer through free-space optics. 
Since there is no need for direct physical contact between the chip and the optical setup operating it, the sensing system can be significantly simplified in a practically preferred form.\cite{wienhold2015all} The detection of the specific molecules has been recently demonstrated on a chip integrated with the fluidic channel based on this approach, where polymer microcavities doped with laser dyes were used for the active resonator.\cite{wienhold2015efficient, wondimu2018robust}
However, in the previous work, the organic laser dye required a pulsed laser as a pump source and had an underlying problem of photobleaching, namely the degradation of photoluminescence.\cite{wienhold2015efficient, wondimu2018robust}
In addition, the demonstrated sensitivity was lower than the usual value generally expected for the WGM resonator sensors.\cite{vollmer2008whispering, he2011detecting, armani2007label, lu2011high, su2016label}
To completely address these issues, we introduce high-sensitive on-chip WGM microcavity sensors embedded with silicon nanoclusters as a stable active compound. Large absorption and emission cross-section of silicon nanoclusters allow pumping and proving remotely through free-space optics, \cite{lupi2011high, kim2016luminescent} which guarantees simple integration with microfluidic channels. 
It also permits the use of an LED as a pumping source for the first time, which significantly reduce the complexity and cost of the measurement setup. 
The nano-gap structure where the mode field is tightly confined and detection events occur is placed in the cavity to increase the sensitivity. 
The molecular detection with this active WGM sensor is demonstrated with the streptavidin-biotin complex, which shows that device sensitivity is 0.012 nm / nM, which is over 20 times larger than that of the previous WGM sensors with remote readout.

\section{Results}

\begin{figure}[t!]
	\begin{center}
		\includegraphics[width=0.95\columnwidth]{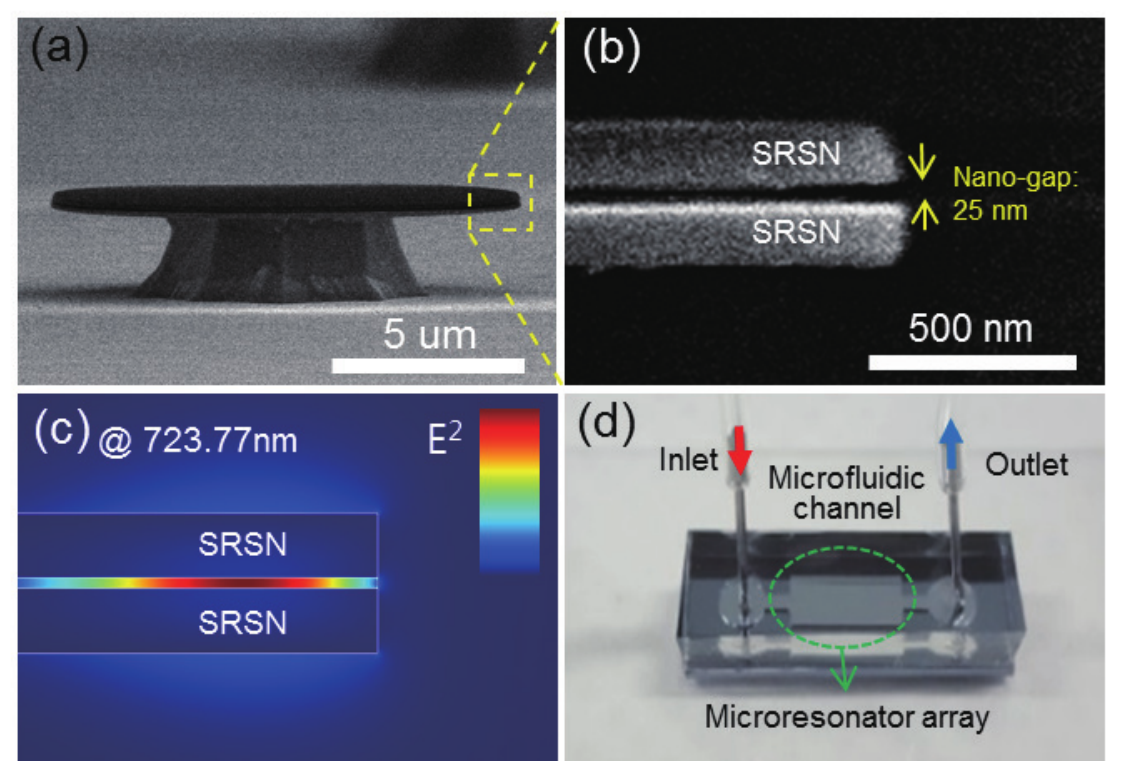}
	\end{center}
	\caption{Images of the microresonator and on-chip sensor (a) SEM image of a disk resonator. (b) A magnified image of the yellow box region in Fig. 1a shows two separate disks and a 25 nm gap. (c) The intensity profile of transverse magnetic (TM) fundamental mode numerically analyzed by COMSOL multi-physics in an aqueous environment. (d) Photograph of the on-chip sensor. The fabricated disk resonators integrated with a microfluidic channel.}
	\label{fig:resonator}
\end{figure}

\subsection{Device development}
The performance enhancement of the developed sensor is achieved by introducing two nanostructures to WGM resonators, that are silicon nanoclusters and nano-gap. 
Since silicon nanoclusters have extremely large absorption cross-section,\cite{park2001band, nguyen2012direct, park2001quantum} they can be pumped not necessarily through resonant modes but direct illumination from the top of the resonator, where interaction length between the pump beam and the resonator structure is only few hundred micrometers. 
The WGM are excited by nanoclusters emitting strong photoluminescence (PL) in the 700 nm wavelength range where absorption by water stays low. 
To form silicon nanoclusters uniformly in a resonator, silicon-rich silicon nitride (SRSN) film is deposited on a substrate and fabricated into disk resonators. 
The excess silicon changes into silicon nanoclusters by a post-annealing process performed in an argon environment. 
The fabrication conditions including excess silicon concentration and annealing temperature are optimized to obtain strong WGM mode in an aqueous environment by the top pumping scheme.   

A nano-gap structure is also introduced in the resonator geometry by placing two pieces of silicon nitride disk plates parallel in the nanometer-scale distance to enhance device sensitivity. 
During the sensing experiment, the gap is filled with Dulbecco's phosphate-buffered saline (DPBS) having a lower refractive index (n=1.33) than that of silicon nitride (n=2) disk plates. 
Due to the requirement for the continuity of the electric displacement field at the interface, this large refractive index contrast in nanometer-scale dimension results in strong confinement of the field intensity in it, which enhances light-matter interaction.\cite{almeida2004guiding, scullion2013slotted}

The fabricated SRSN disk resonators having 12 um diameter and a 120 nm thickness are shown in the scanning electron microscope (SEM) image (Fig. 1a). The gap of 25 nm is clearly shown on the magnified image in Fig. 1b. 
By numerical analysis with COMSOL multi-physics, it is confirmed that $8.6\%$ of the total energy of the fundamental mode is confined in this gap in a water environment as shown in the inset of Fig. 1c, which results in a 6.5-fold enhancement of the sensitivity selectively for the events occurred in the gap.
The fabricated resonator arrays are integrated into a microfluidic channel, which forms a compact optical sensor, as shown in Fig. 1d. 
Microfluidic channels are made of polydimethylsiloxane (PDMS) by conventional soft-lithography. 
The top and side walls of the channels are post-cured to have smooth surfaces through which pump and signal beams pass without excess scattering. 
The analyte solution is injected into the inlet by a syringe pump and guided to the disk array located in the middle of the channel.
A WGM resonator in this encapsulated sensor unit is remotely accessed by a simple experimental setup shown in Fig. 2a and 4a, which facilitates quick and easy operation of the whole sensing process including initial optical alignment. 
The fabrication details for the resonators and channels are described in the experimental section.

\subsection{Optical characteristics}
The optical characteristics of the developed WGM sensors are analyzed in air and aqueous condition. 
As shown in Fig. 2a, the pump beam from a 457.9 nm wavelength argon laser is focused on the top of an SRSN resonator in the fluidic channel. 
The laser intensity on the focal plane is 20 $\si{\watt/\centi\meter^{2}}$ which is enough to excite silicon nanoclusters and get a strong PL emission displaying characteristic cavity resonances. 
The PL emission is collected through the sidewall of the channel by an objective lens having a numerical aperture of 0.15, from which the transverse magnetic (TM) polarization component is selectively captured by a polarizer. 
Since TM modes can be confined in a slot structure more tightly than TE modes, the measurements are performed selectively with the cavity TM modes. 
Fig. 2b and 2c show PL spectrum of TM modes which are measured in air and water environments, respectively. Among many resonant modes shown in Fig. 2b, the fundamental mode family having an intensity profile shown in Fig. 1a is used for sensing. 
The fundamental mode family marked by blue dots is identified by comparing the measured free spectral range (FSR) of each mode family to the value estimated by numerical analysis. 
The FSR of fundamental modes is 7.9 nm, and the quality factor is expected to be larger than 15,000 calculated from the measured full width at half maximum (FWHM) of PL peaks, 0.05 nm, which is limited by the resolution of the spectrometer. 
It is noteworthy that the higher-order modes disappear from the PL spectrum measured in an aqueous environment shown in Fig. 2c. 
The resonator diameter is carefully determined in order that the decreased refractive index contrast between the resonator structure and environment causes serious additional radiation loss selectively to the higher-order modes but not to the fundamental modes. 
This clarified spectrum makes analysis easier by preventing overlaps among different modes. 
The FSR of the modes in Fig. 2c is 7.2 nm which corresponds well to that of the fundamental modes in an aqueous environment estimated by numerical analysis. 
The quality factor is around 8,500 calculated from the FWHM of PL peaks, 0.08 nm.

\begin{figure}[t!]
	\begin{center}
		\includegraphics[width=0.95\columnwidth]{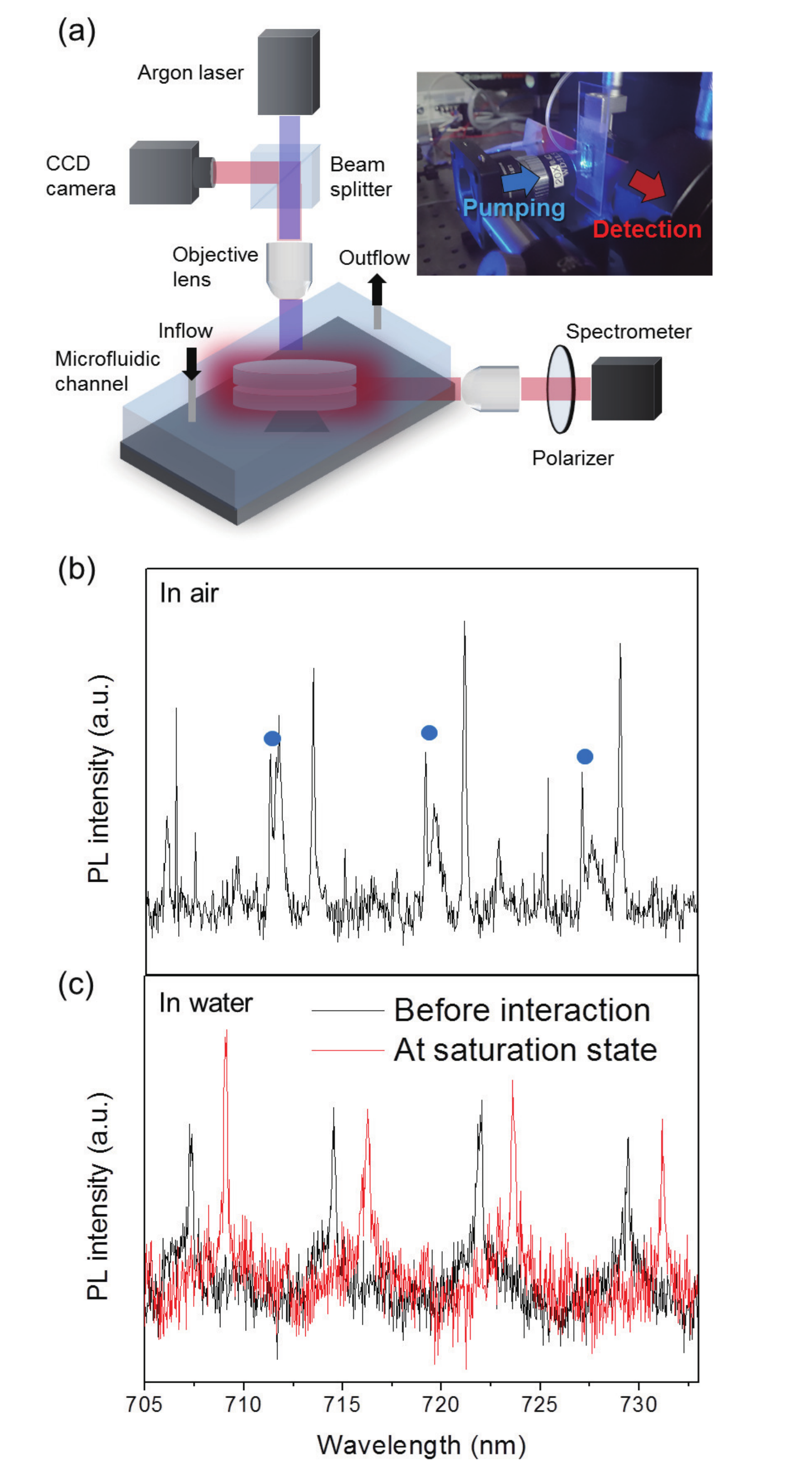}
	\end{center}
	\caption{Experimental setup and measurement results. (a) The schematic of the experimental setup and the photograph of the SRSN disk resonator in a microfluidic channel pumped by the argon laser beam from the top. (b) PL spectrum of an SRSN disk resonator measured in an air environment. Blue circles represent the fundamental TM modes. (c) PL spectrum of biotinylated SRSN disk resonator before (black line) and after (red line) interaction with streptavidin in an aqueous environment.}
	\label{fig2}
\end{figure}

\subsection{The sensitivity of the developed sensor}
The sensitivity of the sensor is verified by the peak-shift measurement for the streptavidin and biotin complex.\cite{liu2017synthetic}
 Before the experiment, SRSN resonators are pretreated with biotin, for which the details can be found in the experimental section.  When the streptavidin in DPBS flows through the fluidic channel, the streptavidin begins to attach to the biotin on the resonator surface, which causes resonance frequency shift. 
 The PL spectra before and after applying streptavidin of 144 nM concentration are displayed as black and red lines in Fig. 2c, respectively. 
 Since the red spectrum is measured after the saturation of the interaction, the relative peak shift of 1.7 nm in the figure gives the sensitivity of our device 0.012 nm/nM. 
 Considering the resolution limit defined by the FWHM of the resonant mode, 0.08 nm, the detection limit of the device is proven to be 6.7 nM for streptavidin. 
 The measured sensitivity of our device is 22 times higher than that of the previously reported active WGM sensor with remote readout.\cite{wondimu2018robust}
 As a control experiment, bovine serum albumin (BSA) also flows into the microfluidic channel with a biotin-pretreated WGM resonator for 35 minutes. 
 Since BSA does not specifically bind to biotin, the BSA induces a much lower frequency shift of 0.25 nm at 1 mg/ml which is a several hundred times higher concentration than streptavidin’s used in the experiments.

\begin{figure}[t!]
	\begin{center}
		\includegraphics[width=0.95\columnwidth]{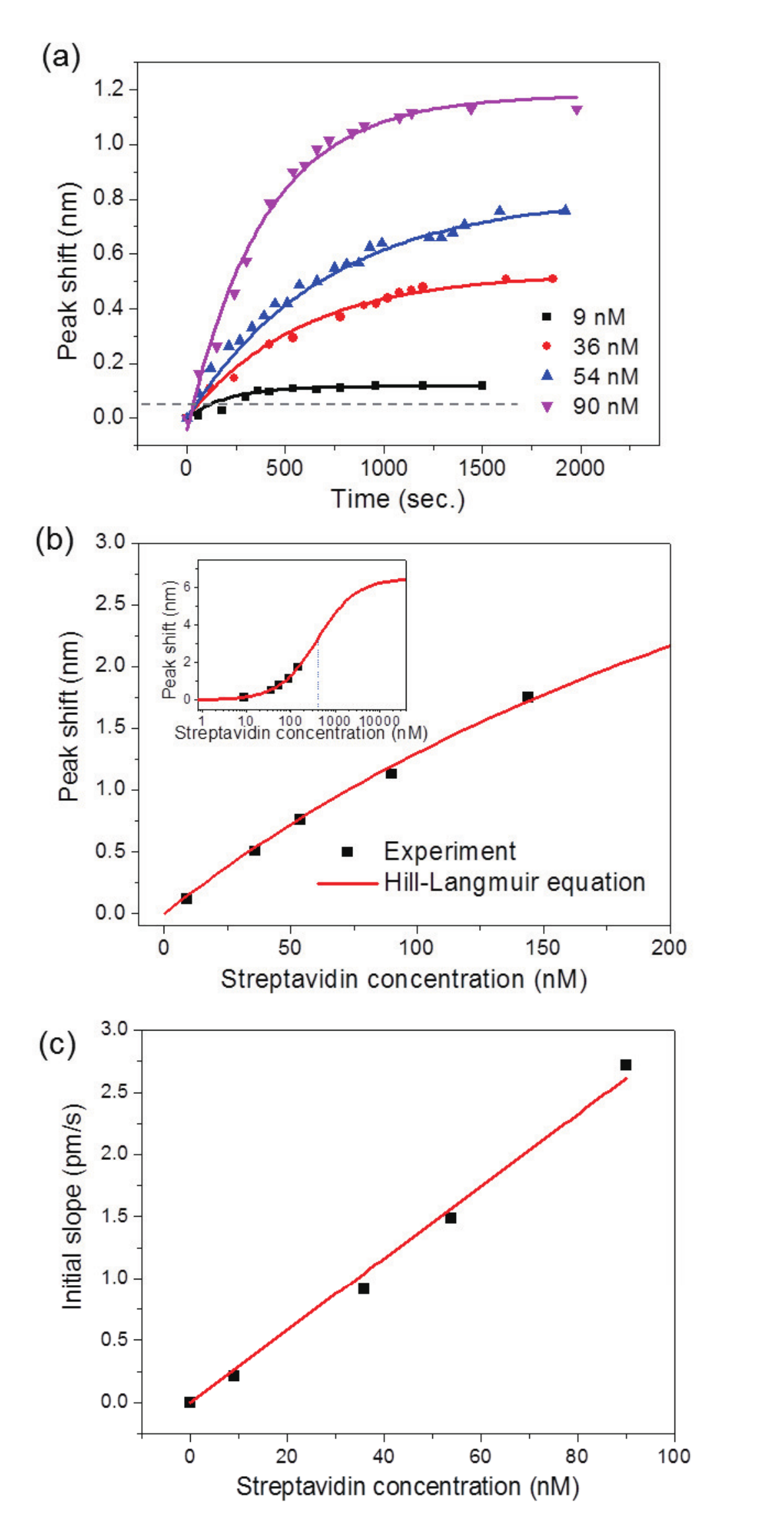}
	\end{center}
	\caption{Biosensing demonstration and results analysis. (a) Real-time measurement of the peak shift for the different concentrations of streptavidin (point). The experimental data are fitted by the association curves based on the Langmuir model (line). The gray dashed line represents the resolution limit of the spectrometer, 0.05 nm. (b) Resonance peak shift in an equilibrium state (black square point) and the results of fitting with Hill-Langmuir equation (red line). The inset of Fig. 3b is in a logarithm scale of Fig. 3b. The blue dashed line represents dissociation constant, $K_{\text {d}}$. (c) The initial slope of the association curve for each concentration in Fig.3a and the results of linear fitting (red line).}
	\label{fig3}
\end{figure}

\subsection{Analysis of Kinetics of streptavidin and biotin complex}
As further investigation for the molecular dynamics and the influence of the nano-gap structure on it, time-transient characteristics of the sensor is measured and analyzed for the streptavidin and biotin complex. 
Fig. 3a shows the relative resonance peak wavelengths which are recorded in time intervals of a few minutes after the injection of streptavidin for four different concentration, 9, 36, 54, 90 nM until the reaction reaches the saturation state. 
These time-transient behaviors can be understood by the Langmuir model, which has been commonly used to analyze the kinetics of surface binding reactions for systems having the ligand immobilized on a solid surface.\cite{halperin2006hybridization,washburn2009label, latour2015langmuir, lowe2017field, hanaor2014scalable, langmuir1918adsorption}
In the model, the response signal of the sensor for the time-dependent molecular reaction is expressed in the form of a characteristic exponential function,\cite{morton1995interpreting} namely association curve, which is displayed as solid lines in Fig. 3a. 
Since the measured data points follow well the trend expected by Langmuir model, the dynamics can be quantitatively analyzed by major parameters, association rate  $k_{\text {on}}$ and dissociation constant $K_{\text {d}}$, of the model. 	

The characteristics of the sensor in the equilibrium state can be understood by the value of $K_{\text {d}}$ and $k_{\text {on}}$ attained from the signals in the saturation regime. 
The Fig. 3b. shows the accumulated resonance shifts according to the concentrations, where each point corresponds to the last point of the serial measurements in Fig. 3a.
These accumulated shifts are fitted to the Hill-Langmuir equation represented by the red line. 
The inset of Fig. 3b, which displays the same data in logarithmic scale, clearly shows the value $K_{\text {d}}$ of $ 3.8 \times 10^{-7}\si{\molar}$ as the blue dashed line where the slope of the curve is the steepest. 
By applying this $K_{\text {d}}$ to the equation for the association curves in Fig. 3a, the $k_{\text {on}}$ values can be obtained. 
The measured $k_{\text {on}}$ is  $1.2 \times 10^3 \si{\per\molar\per\second}$, $4.2 \times 10^3 \si{\per\molar\per\second}$, $3.2 \times 10^3 \si{\per\molar\per\second}$ and $6.4 \times 10^3 \si{\per\molar\per\second}$  for the streptavidin concentrations of 9 nM, 36 nM, 54 nM and 90 nM, respectively. 
The slightly larger $k_{\text {on}}$ of 9 nM streptavidin is caused by the detection limit of the spectrometer, 0.05 nm, which is marked as a dashed line in Fig. 3a. 
This detection limit is not enough to finely resolve the kinetics in this low concentration. 
These measured values of $k_{\text {on}}$ and $K_{\text {d}}$ are similar to the previously reported results based on the sensors having the geometry open to the sensing medium.\cite{wondimu2018robust, nair2017crossed, d2008real}
Thus, we conclude that the 25 nm gap, which enhance the sensitivity, does not significantly affect on molecular dynamics in the equilibrium state.

The characteristics in the initial state which are affected by the diffusion of the molecules from the medium into the nano-gap are also quantitatively assessed by $k_{\text {on}}$. 
Under the mass transport limitation commonly occurring in the molecular kinetic analysis, the initial slope of the association curve is linearly proportional to the analyte concentration by Fick’s first law.\cite{washburn2009label, eddowes1987direct}
The initial slopes of the association curves in Fig. 3a are displayed as the black point in the Fig.3c, where the points are linearly aligned as expected by Fick’s law. 
From the slope of this red line, we can obtain $k_{\text {on}}$ of $4.8 \times 10^3 \si{\per\molar\per\second}$, which corresponds well to $4.6 \times 10^3 \si{\per\molar\per\second}$, the average value of the $k_{\text {on}}$ previously obtained for three different concentrations except for 9 nM. 
This means that there is no considerable lag of the sensor response due to the diffusion into the nano-gap. 
In other word, the initial concentration of streptavidin in the nano-gap reaches quickly to that of the sensing medium in a few minuites, the time resolution of this measurement. 

\begin{figure}[t!]
	\begin{center}
		\includegraphics[width=0.95\columnwidth]{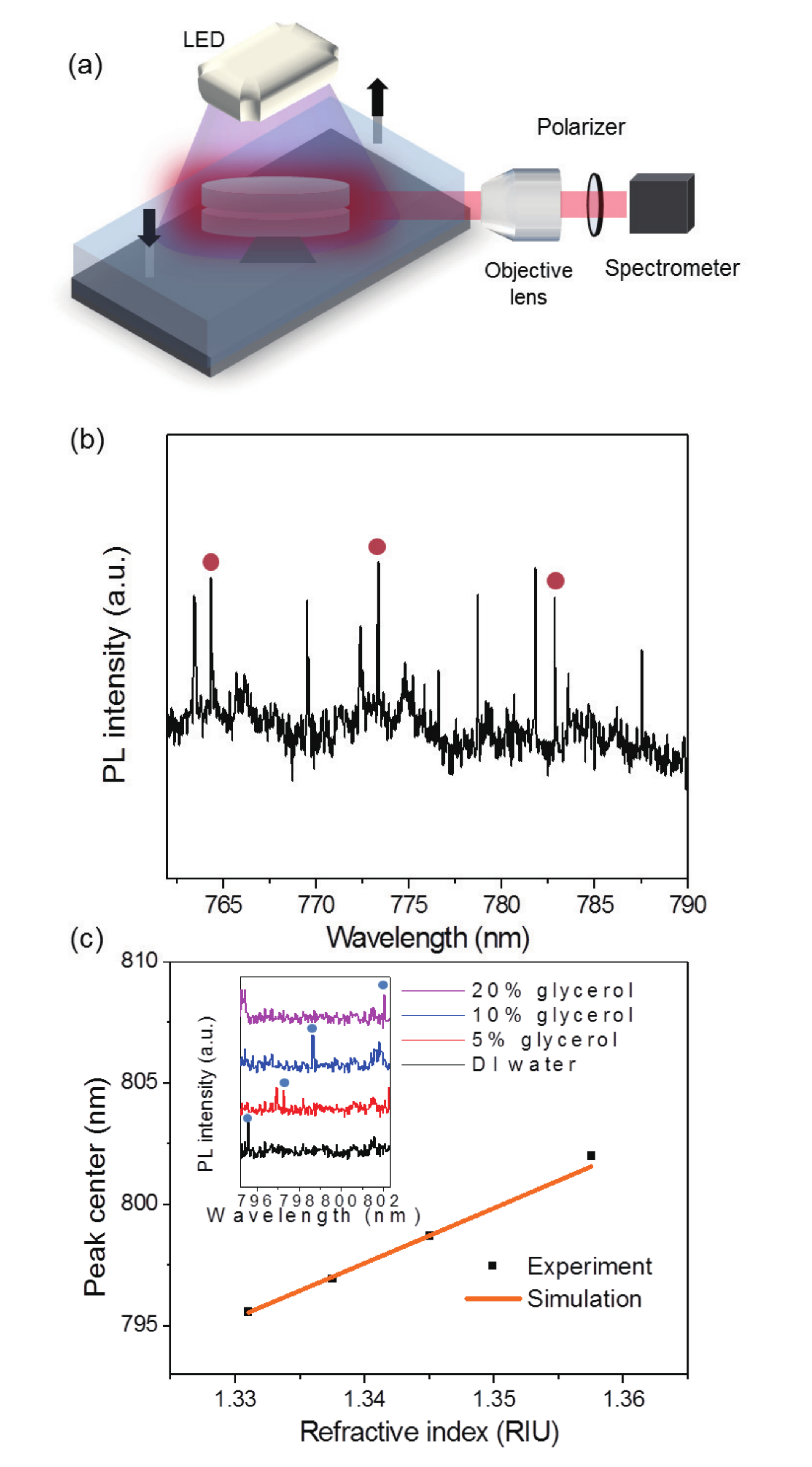}
	\end{center}
	\caption{The measurement with a LED light source and refractometric sensing. (a) The schematic of the measurement setup based on an LED pump used in this experiment. (b) PL spectrum of an SRSN microresonator induced by a single LED pump irradiation in an air environment. The red circles represent the fundamental TM mode. (c) The response of the sensor to the change of the refractive index of the medium. The inset of Fig. 4c shows the resonance peak shift induced by the injection of different refractive index solutions of glycerol diluted in DI water through the microfluidic channel. The black square points indicate the peak shift of resonance wavelength and the orange line represents the numerically analyzed sensitivity of the resonator.}
	\label{fig4}
\end{figure}

\subsection{On-chip sensor platform with LED pump source} 
The measurement setup can be simplified further by adopting an LED as a pump source. 
Due to the high absorption cross-section of silicon nanoclusters, argon laser, the pump source of SRSN resonators, can be replaced by commercialized high-intensity LEDs which emit light in the absorption wavelength of silicon nanoclusters ranging from under 200 nm to 470 nm.\cite{deshpande1995optical}
Since the light-emitting area of LED is much larger than the resonator, simply locating the microfluidic chip under the LED module is enough for the alignment. 
Therefore, there is no need for the pump laser and additional components such as a beam splitter, objective lens, and CCD which are required for the focusing and precise alignment of the laser beam. 
Since the intensity of commercially available LEDs is lower than that of Ar laser and silicon nanoclusters absorb UV more efficiently than the visible,\cite{lupi2011high} the LED with a center wavelength of 365 nm (LUMINUS SST10-UV) is used as the top pump source. 
The effective intensity of the LED incident on the resonator chip surface which is 4 mm apart from the LED is approximately 5.6 $\si{\watt/\centi\meter^{2}}$ with ignoring the absorption by the PDMS wall of the fluidic channel. 
Although the intensity of LED is far less than the argon laser (20 $\si{\watt/\centi\meter^{2}}$), it is sufficient to excite WGM of the resonators in an aqueous environment and results in resonant peaks of the PL emission spectrum as shown in Fig. 4b. 
The FSR of the TM fundamental modes and the quality factor are exactly the same to the values with the argon pump laser, namely 9.1 nm and 15000, respectively.
To verify the proper operation of the LED pumping scheme, we perform on-chip bulk sensing instead of specific biomolecule detection because ultraviolet light from an LED deforms biomolecules. 
In bulk sensing, the concentration of the target chemicals homogeneously dispersed in the medium can be measured by detecting the change of the refractive index. 
The glycerol is diluted in deionized water in three different concentrations: 5, 10, and 20 $\%$, which resulted in the increase of the refractive index of the mixed mediums from that of DI water, 1.331, to 1.338, 1.345, and 1.359, respectively.\cite{hoyt1934new}
The inset of Fig. 4c shows the frequency of the resonant peak for the serial injections of the diluted glycerol in three different concentrations. 
It closely matches the linear frequency shift of 226.67 nm /refractive index unit (RIU) numerically estimated by COMSOL Multiphysics for a TM fundamental mode which is plotted in the orange line in Fig. 4c. 
Considering the resolution limit of 0.08 nm defined by the FWHM of the resonant mode, the detection limit for the change of the glycerol concentration is around 0.25 $\%$. 
We expect that the detection of biomolecules is possible based on the developed sensors with an LED pump if the intensity of LEDs in the visible wavelength increases to around 20 $\si{\watt/\centi\meter^{2}}$ or the absorption and emission efficiency of silicon nanoclusters are further improved.

\section{Experimental}

\subsection{Fabrication of SRSN disk resonator with nano-gap}
The fabrication process for the actual devices starts with thin film deposition of SRSN, silicon dioxide, SRSN, and amorphous silicon on the silicon substrate by ion beam sputtering. 
After deposition, the disks of 12 ㎛ diameter is patterned on the multilayer film by conventional photolithography. 
Then, amorphous silicon, the top of the multilayer, is dry-etched by inductively coupled plasma-reactive ion etching (ICP-RIE) in CHF3 and SF6 gas environments and used as a hard mask for following multilayer etching. 
The whole multilayer of SRSN, silicon dioxide, and SRSN is dry-etched simultaneously by ICP-RIE, in CHF3 and O2 gas environments. 
To remove the amorphous silicon used as a hard mask and form a silicon post at the bottom of the resonator, the device is wet-etched in a 60 ℃ potassium hydroxide solution. 
The device is annealed in an argon gas environment at 1100 ℃ for 30 minutes to form silicon nanoclusters in the SRSN film. Finally, the silicon dioxide layer is wet-etched to form a nano-gap between the SRSN disks with a depth of 1 ㎛ by a buffered oxide etchant.

\subsection{Microfluidic channel fabrication}
A channel master having a height of 270 um, a width of 3 mm, and a length of 8 mm is formed on a silicon substrate by SU-8 photoresist using a conventional soft lithography process.\cite{abdelgawad2008soft} 
A PDMS mixture (PDMS base and curing agent) is poured on the SU-8 channel master and degassed in a vacuum chamber to remove bubbles from the PDMS mixture and left it in the oven at 65 ℃ for 3 hours to solidify. 
The solid PDMS is stripped from the channel master, cut into one channel unit and punched for connection of the inlet and outlet tubes. 
The side surface of the microfluidic channel through which the light emission of the disk resonator passes is applied by a PDMS mixture and cured in the oven once more to smooth the rough surface formed when the channel is cut by blazes. 
The PDMS thicknesses of the window in the microfluidic channel through which the pump light and the emission light from the resonator passes are about 1.7 mm and 1.5 mm, respectively. 
To integrate the microfluidic channels with the disk resonator substrate, both sides are in contact with each other after oxygen plasma treatment for 60 seconds.\cite{scullion2013slotted} 

\subsection{Biotin immobilization on the disk surface} 
For real-time measurement of the streptavidin and biotin interaction, an SRSN disk resonator surface is pretreated with biotin.\cite{williams2012immobilization} 
The fabricated device surface is immersed in an RCA solution (H2O: H2O2: NH3 = 5:1:1) at 65℃ for 10 minutes to clean and expose the OH group on the surface. 
Then the substrate is treated by oxygen plasma for 60 seconds to integrate the microfluidic channel with the resonator substrate. 
After the microfluidic channel is integrated into the resonator substrate, 5$\%$ APTES (3-aminopropyltriethoxysilane, Sigma-Aldrich) in absolute ethanol (EMSURE®) is injected into the microfluidic channel, then left for 2 hours to form the amine-group on the resonator surface. 
The channel is then flushed several times with ethanol to wash out the unreacted APTES from the device and finally rinsed with deionized water, and the device is placed on the hot plate at 100 ℃ for 30 minutes in order to cure the APTES. 
Finally, 100 µg biotin N-hydroxysuccinimide ester (Sigma-Aldrich) in 1 ml DPBS is flowed into the microfluidic channel and left for 20 hours at room temperature to combine biotin N-hydroxysuccinimide ester with amine-groups on the surface. 

\section{Conclusions}
We developed label-free sensors with active WGM resonators integrated with microfluidic channels in a practical form which is free from the strict evanescent coupling scheme relying on tapered fiber. 
The sensitivity of this device confirmed by real-time measurements for the streptavidin-biotin complex is 0.012 nm/nM, which is more than 20 times larger than that of the previously reported active WGM sensors without direct physical coupling. Since the characteristic parameters such as $k_{\text {on}}$ and $K_{\text {d}}$ attained by analyzing the dynamics of the streptavidin-biotin complex agree well with the previously reported values, it is concluded that the nano-slot structure having 25 nm gap introduced to enhance the sensitivity doesn’t distort the molecular dynamics significantly. 
Furthermore, by taking advantage of the large absorption cross-section of silicon nanoclusters, WGM sensing based on the direct illumination of an LED pump is demonstrated for the first time. 
We expect that this approach paves the way to implement cost-effective on-chip label-free sensors for practical applications which can be operated in a simple measurement setup for ease of use.

In addition, the nano-gap structure where molecular interactions take place can be applied for the observation of the real-time molecular dynamics in nanochannels having a dimension comparable to the molecular complex. 
Considering the size of streptavidin and biotin (around 5 nm and 1 nm, respectively) and the fact that they attach on both surfaces in the slot, a slight reduction of the gap width from the current design is expected to induce observable changes of the dynamics.

\bibliographystyle{unsrt}
\bibliography{ref}

\begin{thebibliography}{10}

\bibitem{Eugene2017towards}
E.~Kim, M.~D. Baaske, and F.~Vollmer.
\newblock Towards next-generation label-free biosensors: recent advances in
  whispering gallery mode sensors.
\newblock {\em Lab on a Chip}, 17(7):1190--1205, 2017.

\bibitem{vollmer2008whispering}
F.~Vollmer and S.~Arnold.
\newblock Whispering-gallery-mode biosensing: label-free detection down to
  single molecules.
\newblock {\em Nature methods}, 5(7):591, 2008.

\bibitem{vollmer2012review}
F.~Vollmer and L.~Yang.
\newblock Review label-free detection with high-q microcavities: a review of
  biosensing mechanisms for integrated devices.
\newblock {\em Nanophotonics}, 1(3-4):267--291, 2012.

\bibitem{subramanian2018label}
S.~Subramanian, H.-Y. Wu, T.~Constant, J.~Xavier, and F.~Vollmer.
\newblock Label-free optical single-molecule micro-and nanosensors.
\newblock {\em Advanced Materials}, 30(51):1801246, 2018.

\bibitem{baaske2014single}
Martin~D Baaske, Matthew~R Foreman, and Frank Vollmer.
\newblock Single-molecule nucleic acid interactions monitored on a label-free
  microcavity biosensor platform.
\newblock {\em Nature nanotechnology}, 9(11):933, 2014.

\bibitem{he2011detecting}
L.~He, {\c{S}}.~K. {\"O}zdemir, J.~Zhu, W.~Kim, and L.~Yang.
\newblock Detecting single viruses and nanoparticles using whispering gallery
  microlasers.
\newblock {\em Nature nanotechnology}, 6(7):428, 2011.

\bibitem{armani2007label}
A.~M. Armani, R.~P. Kulkarni, S.~E. Fraser, R.~C. Flagan, and K.~J. Vahala.
\newblock Label-free, single-molecule detection with optical microcavities.
\newblock {\em science}, 317(5839):783--787, 2007.

\bibitem{zhu2010chip}
Jiangang Zhu, Sahin~Kaya Ozdemir, Yun-Feng Xiao, Lin Li, Lina He, Da-Ren Chen,
  and Lan Yang.
\newblock On-chip single nanoparticle detection and sizing by mode splitting in
  an ultrahigh-q microresonator.
\newblock {\em Nature photonics}, 4(1):46, 2010.

\bibitem{arnold2010whispering}
S.~Arnold, S.~I. Shopova, and S.~Holler.
\newblock Whispering gallery mode bio-sensor for label-free detection of single
  molecules: thermo-optic vs. reactive mechanism.
\newblock {\em Optics Express}, 18(1):281--287, 2010.

\bibitem{foreman2015whispering}
Matthew~R Foreman, Jon~D Swaim, and Frank Vollmer.
\newblock Whispering gallery mode sensors.
\newblock {\em Advances in optics and photonics}, 7(2):168--240, 2015.

\bibitem{wienhold2015all}
T.~Wienhold, S.~Kraemmer, S.~F. Wondimu, T.~Siegle, U.~Bog, U.~Weinzierl,
  S.~Schmidt, H.~Becker, H.~Kalt, T.~Mappes, S.~Koeberaf, and C.~Koos.
\newblock All-polymer photonic sensing platform based on whispering-gallery
  mode microgoblet lasers.
\newblock {\em Lab on a Chip}, 15(18):3800--3806, 2015.

\bibitem{xu2018wireless}
X.~Xu, W.~Chen, G.~Zhao, Y.~Li, C.~Lu, and L.~Yang.
\newblock Wireless whispering-gallery-mode sensor for thermal sensing and
  aerial mapping.
\newblock {\em Light: Science \& Applications}, 7(1):62, 2018.

\bibitem{knight1997phase}
J.~C. Knight, G.~Cheung, F.~Jacques, and T.~A. Birks.
\newblock Phase-matched excitation of whispering-gallery-mode resonances by a
  fiber taper.
\newblock {\em Optics letters}, 22(15):1129--1131, 1997.

\bibitem{barrios2012integrated}
C.~A. Barrios.
\newblock Integrated microring resonator sensor arrays for labs-on-chips.
\newblock {\em Analytical and bioanalytical chemistry}, 403(6):1467--1475,
  2012.

\bibitem{wienhold2015efficient}
T.~Wienhold, S.~Kraemmer, A.~Bacher, H.~Kalt, C.~Koos, S.~Koeber, and
  T.~Mappes.
\newblock Efficient free-space read-out of wgm lasers using circular
  micromirrors.
\newblock {\em Optics express}, 23(2):1025--1034, 2015.

\bibitem{wondimu2018robust}
S.~F. Wondimu, M.~Hippler, C.~Hussal, A.~Hofmann, S.~Kr{"a}mmer, J.~Lahann,
  H.~Kalt, W.~Freude, and C.~Koos.
\newblock Robust label-free biosensing using microdisk laser arrays with
  on-chip references.
\newblock {\em Optics express}, 26(3):3161--3173, 2018.

\bibitem{lu2011high}
T.~Lu, H.~Lee, T.~Chen, S.~Herchak, J.-H. Kim, S.~E. Fraser, R.~C. Flagan, and
  K.~J. Vahala.
\newblock High sensitivity nanoparticle detection using optical microcavities.
\newblock {\em Proceedings of the National Academy of Sciences},
  108(15):5976--5979, 2011.

\bibitem{su2016label}
J.~Su, A.~FG Goldberg, and B.~M. Stoltz.
\newblock Label-free detection of single nanoparticles and biological molecules
  using microtoroid optical resonators.
\newblock {\em Light: Science \& Applications}, 5(1):e16001, 2016.

\bibitem{lupi2011high}
F.~F. Lupi, D.~Navarro-Urrios, J.~Monserrat, C.~Dominguez, P.~Pellegrino, and
  B.~Garrido.
\newblock High q light-emitting si-rich si 3 n 4 microdisks.
\newblock {\em Optics Letters}, 36(8):1344--1346, 2011.

\bibitem{kim2016luminescent}
G.~Kim and J.~H. Shin.
\newblock Luminescent silicon-rich nitride horizontal air-slot microdisk
  resonators for biosensing.
\newblock {\em IEEE Photonics Technology Letters}, 28(21):2331--2334, 2016.

\bibitem{park2001band}
Nae-Man Park, Tae-Soo Kim, and Seong-Ju Park.
\newblock Band gap engineering of amorphous silicon quantum dots for
  light-emitting diodes.
\newblock {\em Applied Physics Letters}, 78(17):2575--2577, 2001.

\bibitem{nguyen2012direct}
PD~Nguyen, DM~Kepaptsoglou, QM~Ramasse, and A~Olsen.
\newblock Direct observation of quantum confinement of si nanocrystals in
  si-rich nitrides.
\newblock {\em Physical Review B}, 85(8):085315, 2012.

\bibitem{park2001quantum}
Nae-Man Park, Chel-Jong Choi, Tae-Yeon Seong, and Seong-Ju Park.
\newblock Quantum confinement in amorphous silicon quantum dots embedded in
  silicon nitride.
\newblock {\em Physical review letters}, 86(7):1355, 2001.

\bibitem{almeida2004guiding}
V.~R. Almeida, Q.~Xu, C.~A. Barrios, and M.~Lipson.
\newblock Guiding and confining light in void nanostructure.
\newblock {\em Optics letters}, 29(11):1209--1211, 2004.

\bibitem{scullion2013slotted}
M.~Scullion, T.~Krauss, and A.~Di~Falco.
\newblock Slotted photonic crystal sensors.
\newblock {\em Sensors}, 13(3):3675--3710, 2013.

\bibitem{liu2017synthetic}
W.~Liu, S.~K. Samanta, B.~D. Smith, and L.~Isaacs.
\newblock Synthetic mimics of biotin/(strept) avidin.
\newblock {\em Chemical Society Reviews}, 46(9):2391--2403, 2017.

\bibitem{halperin2006hybridization}
A.~Halperin, A.~Buhot, and E.~B. Zhulina.
\newblock On the hybridization isotherms of dna microarrays: the langmuir model
  and its extensions.
\newblock {\em Journal of Physics: Condensed Matter}, 18(18):S463, 2006.

\bibitem{washburn2009label}
A.~L. Washburn, L.~C. Gunn, and R.~C. Bailey.
\newblock Label-free quantitation of a cancer biomarker in complex media using
  silicon photonic microring resonators.
\newblock {\em Analytical chemistry}, 81(22):9499--9506, 2009.

\bibitem{latour2015langmuir}
R.~A. Latour.
\newblock The langmuir isotherm: a commonly applied but misleading approach for
  the analysis of protein adsorption behavior.
\newblock {\em Journal of Biomedical Materials Research Part A},
  103(3):949--958, 2015.

\bibitem{lowe2017field}
B.~M. Lowe, K.~Sun, I.~Zeimpekis, C.-K. Skylaris, and N.~G. Green.
\newblock Field-effect sensors--from ph sensing to biosensing: sensitivity
  enhancement using streptavidin--biotin as a model system.
\newblock {\em Analyst}, 142(22):4173--4200, 2017.

\bibitem{hanaor2014scalable}
D.~A. Hanaor, M.~Ghadiri, W.~Chrzanowski, and Y.~Gan.
\newblock Scalable surface area characterization by electrokinetic analysis of
  complex anion adsorption.
\newblock {\em Langmuir}, 30(50):15143--15152, 2014.

\bibitem{langmuir1918adsorption}
Irving Langmuir.
\newblock The adsorption of gases on plane surfaces of glass, mica and
  platinum.
\newblock {\em Journal of the American Chemical society}, 40(9):1361--1403,
  1918.

\bibitem{morton1995interpreting}
Thomas~A Morton, David~G Myszka, and Irwin~M Chaiken.
\newblock Interpreting complex binding kinetics from optical biosensors: a
  comparison of analysis by linearization, the integrated rate equation, and
  numerical integration.
\newblock {\em Analytical biochemistry}, 227(1):176--185, 1995.

\bibitem{nair2017crossed}
Srijit Nair, Carlos Escobedo, and Ribal~Georges Sabat.
\newblock Crossed surface relief gratings as nanoplasmonic biosensors.
\newblock {\em ACS sensors}, 2(3):379--385, 2017.

\bibitem{d2008real}
R.~D’Agata, G.~Grasso, and G.~Spoto.
\newblock Real-time binding kinetics monitored with surface plasmon resonance
  imaging in a diffusion-free environment.
\newblock {\em Open Spectrosc J}, 2:1--9, 2008.

\bibitem{eddowes1987direct}
M.~J. Eddowes.
\newblock Direct immunochemical sensing: basic chemical principles and
  fundamental limitations.
\newblock {\em Biosensors}, 3(1):1--15, 1987.

\bibitem{deshpande1995optical}
S.~V. Deshpande, E.~Gulari, S.~W. Brown, and S.~C. Rand.
\newblock Optical properties of silicon nitride films deposited by hot filament
  chemical vapor deposition.
\newblock {\em Journal of Applied Physics}, 77(12):6534--6541, 1995.

\bibitem{hoyt1934new}
L.~F. Hoyt.
\newblock New table of the refractive index of pure glycerol at 20 c.
\newblock {\em Industrial \& Engineering Chemistry}, 26(3):329--332, 1934.

\bibitem{abdelgawad2008soft}
M.~Abdelgawad, M.~W. Watson, E.~W. Young, J.~M. Mudrik, M.~D. Ungrin, and A.~R.
  Wheeler.
\newblock Soft lithography: masters on demand.
\newblock {\em Lab on a Chip}, 8(8):1379--1385, 2008.

\bibitem{williams2012immobilization}
E.~H. Williams, A.~V. Davydov, A.~Motayed, S.~G. Sundaresan, P.~Bocchini, L.~J.
  Richter, G.~Stan, K.~Steffens, R.~Zangmeister, J.~A. Schreifels, and M.~V.
  Rao.
\newblock Immobilization of streptavidin on 4h--sic for biosensor development.
\newblock {\em Applied surface science}, 258(16):6056--6063, 2012.

\end{thebibliography}

\end{document}